\documentclass[twocolumn,
amsmath,
amssymb,
amsfonts,
final,
a4paper,
aps,
citeautoscript,
footinbib,
prl,
superscriptaddress]{revtex4}
\usepackage{times}
\usepackage{graphicx}
\usepackage[latin1]{inputenc}
\usepackage[
   breaklinks=true,
   colorlinks=true,
   bookmarks=true,
   bookmarksopenlevel=3, 
   bookmarksnumbered=true,
   menucolor=black,
   linkcolor=blue,
   citecolor=blue,
   urlcolor=blue,
   pdftitle={},
   pdfauthor={Guillaume Roux},
   pdfcreator={GhostScript},
   pdfsubject={},
   pdfkeywords={},
   pdfstartview=FitH]{hyperref}

\usepackage{physics}

\begin{document}

\author{Bradraj \surname{Pandey}}
\affiliation{LPTMS, CNRS, Univ. Paris-Sud, Universit\'e Paris-Saclay, 91405 Orsay, France}
\author{Kirill \surname{Plekhanov}}
\affiliation{LPTMS, CNRS, Univ. Paris-Sud, Universit\'e Paris-Saclay, 91405 Orsay, France}
\affiliation{Centre de Physique Th\'eorique, Ecole Polytechnique, CNRS, Universit\'e Paris-Saclay, F-91128 Palaiseau, France}
\author{Guillaume \surname{Roux}}
\email{guillaume.roux@u-psud.fr}
\affiliation{LPTMS, CNRS, Univ. Paris-Sud, Universit\'e Paris-Saclay, 91405 Orsay, France}

\date{February 13$^{\text{th}}$ 2018}
 
\title{Quantum purification spectroscopy}


\begin{abstract}
We discuss a protocol based on quenching a purified quantum system that allows to capture bulk spectral features.
It uses an infinite temperature initial state and an interferometric strategy to access the Loschmidt amplitude, from which the spectral features are retrieved via Fourier transform, providing coarse-grained approximation at finite times.
It involves techniques available in current experimental setups for quantum simulation, at least for small systems.
We illustrate possible applications in testing the eigenstate thermalization hypothesis and the physics of many-body localization.
\end{abstract}

\maketitle 

Throughout the joint recent progresses in experiments and theoretical approaches, new quantities and protocols have been proposed to tackle the quantum many-body problem.
With the rise of quantum simulators~\cite{Georgescu2014}, new perspectives involving sophisticated protocols can be envisioned, which are naturally based on time-evolved observables and probabilities, such as the celebrated Loschmidt echo~\cite{Peres1984,Gorin2006}. A quantum quench, or evolving an arbitrary initial state under the action of a given Hamiltonian, is one of the simplest out-of-equilibrium protocol easily achievable on such platforms. As an example, the field of dynamical quantum phase transitions~\cite{Heyl2013, Andraschko2014a, Zvyagin2017, Polkovnikov2017, Heyl2018a} has emerged, with recent experimental observations~\cite{Jurcevic2017, Flaeschner2018, Zhang2017}, and its generalization to finite temperatures~\cite{Abeling2016,Sedlmayr2017,Mera2018}. This generalization is connected to quantum chaos, the so-called scrambling of information~\cite{Shenker2014,Swingle2016} and the statistics of work done in a quench~\cite{Silva2008, CamposVenuti2010}, via the double thermal field picture~\cite{Campo2017, Campo2017a, Chenu2017}, or purification scheme in quantum information. In parallel, another quantum chaos related observable is the out-of-time-order correlator~\cite{Swingle2016,Zhu2016,Swingle2017}, also recently measured~\cite{Li2017}, and which proves useful in analysing the dynamics of many-body systems~\cite{Yichen2016,Heyl2018,Bordia2018}. 

In this paper, we combine some of these ideas to propose a protocol that aims at performing the spectroscopy of energy, observables, and Fock states of a given Hamiltonian.
We believe it is relevant for current experimental setups with few degrees of freedom, and it also defines numerically tractable quantities for current techniques such a matrix-product states (MPS)~\cite{Schollwoeck2011}. It naturally defines semi-classical, or coarse-grained, quantities that could be compared to semi-classical predictions. We show possible applications in testing the eigenstate thermalization hypothesis (ETH)~\cite{Jancel1969,Peres1984,Deutsch1991,Srednicki1994,Tasaki1998,Srednicki1999,Rigol2008,Reimann2008,Biroli2010,DAlessio2015,Borgonovi2016} and many-body localization (MBL)~\cite{Basko2006,Oganesyan2007,Znidaric2008,Pal2010,Murphy2011,Serbyn2013,Luca2013,Kjaell2014,Luitz2015,Schreiber2015,Nandkishore2015,Altman2015,Alet2018}.

Using standard notations for quantum quenches, we consider an Hamiltonian $H$ leaving on an Hilbert space of dimension $D$, of eigenstates/energies $\{\ket{n},E_n\}_{n=1,D}$.
For a quench starting from the initial state $\ket{\psi_0}$, we define the return probability, or Loschmidt echo, $L(t)=|G(t)|^2$, where the Loschmidt amplitude (the time autocorrelation function~\cite{Tannor2007}) reads
$G(t) = \expval{e^{-it H}}{\psi_0} = \sum_n p_n e^{-itE_n}$,
with weights $p_n = |{\braket{n}{\psi_0}}|^2$.
These weights constitute the diagonal ensemble, or energy distribution set by the initial state.
In the physics of dynamical quantum phase transitions, $G(t)$ is understood as a boundary partition function, extended to the complex plane. 
In what follows, we denote the complex partition function by $Z(z) = \Tr(e^{-z H})= \sum_n e^{-zE_n}$, with $z \in\mathbb{C}$.
When considering some observable $A$, we write eigenstates expectation values as $A_n = \expval{A}{n}$.
The local degrees of freedom at site $\ell$ over a lattice of size $L$ are described by quantum numbers $\sigma_\ell$ and $\{\ket{\sigma}=\ket{\sigma_1,\ldots,\sigma_L}\}$ will be the set of Fock states.
The following discussion does not make restrictive assumptions on $H$, and we illustrate the ideas on the quantum Ising Hamiltonian~\cite{Dutta2015}:
\begin{equation}
\label{eq:HamIsing}
H = -J_z\sum_{\ell=1}^{L-1} S^z_{\ell}S^z_{\ell + 1} -h_x\sum_\ell S^x_\ell -h_z\sum_\ell S^z_\ell\,,
\end{equation}
with $D=2^L$, and $S^\alpha$ the spin-$1/2$ operator in the $\alpha$ direction.

\textbf{A purification protocol to measure $Z(it)$} -- Following the purification representation~\cite{Uhlmann1976,Uhlmann1986,Nielsen2000}, we consider two copies $S$ (system) and $\tilde{S}$ (ancillas) of the same degrees of freedom, that are initially entangled in the infinite temperature state:
\begin{equation*}
\ket{\psi_\infty}
= \frac{1}{\sqrt{D}}\prod_{\ell=1}^L \sum_{\sigma_\ell}\ket{\sigma_\ell}_S \ket{\sigma_\ell}_{\tilde{S}}
= \frac{1}{\sqrt{D}}\sum_{\sigma} \ket{\sigma}_{S}\ket{\sigma}_{\tilde{S}} \;.
\end{equation*}
The reduced density matrix describing $S$ is obtained by tracing out the ancillas.
After decoupling $S$ from $\tilde{S}$, we let the system evolve under $H$ while $\tilde{S}$ is not evolving:
\begin{equation}
\ket{\psi(t)} = \frac{1}{\sqrt{D}}\sum_{\sigma} \left(e^{-it H}\ket{\sigma}_S\right)\otimes\ket{\sigma}_{\tilde{S}}\;.
\end{equation}
At time $t$, projecting the evolving state onto $\ket{\psi_{\infty}}$ yields
\begin{align}
G(t)= \braket{\psi_{\infty}}{\psi(t)} &= \frac{1}{D}\sum_{\sigma,\tilde\sigma} \bra{\tilde\sigma}e^{-it H}\ket{\sigma}_S\times\bra{\tilde\sigma}\ket{\sigma}_{\tilde{S}} \nonumber \\
 &= \frac{1}{D} \Tr(e^{-it H}) = \frac{Z(it)}{D} \;.
\end{align}
The protocol \textit{qualitatively} amounts -- only regarding energy measurements -- to work with a uniformly distributed diagonal ensemble $p_n = 1/D$.

\textbf{Accessing the $G(t)$ function} -- 
Experimentally, we propose the following protocol, having in mind small systems in current realistic setups. First, in order to measure the complex valued overlap $\braket{\psi_0}{\psi(t)}$ between an initial state $\ket{\psi_0}$ and a time-evolved state $\ket{\psi(t)} = e^{-itH}\ket{\psi_0}$, one can use an interferometric technique using a probe qubit $\ket{q}_{\text{p}}$~\cite{Sjoeqvist2000,Viyuela2013,Viyuela2018,Swingle2016}. Starting from $\ket{\psi_0}\ket{0}_{\text{p}}$, one rotates the probe state to $\frac{1}{\sqrt{2}}(\ket{\psi_0}\ket{0}_{\text{p}} + \ket{\psi_0}\ket{1}_{\text{p}})$, and evolves the system through a control-$U$ gate~\cite{Nielsen2000} that applies $e^{-itH}$ conditionally on the $\ket{1}_{\text{p}}$ state. One gets after time $t$ the superposition $\frac{1}{\sqrt{2}}(\ket{\psi_0}\ket{0}_{\text{p}} + \ket{\psi(t)}\ket{1}_{\text{p}})$. Last, one measures the real and imaginary parts of $\braket{\psi_0}{\psi(t)}$ through the expectation values of Pauli matrices $\expval{\sigma_x}_{\text{p}}$ and $\expval{\sigma_y}_{\text{p}}$ computed in the reduced density matrix of the probe. 
The main difficulty lies in the control-$U$ gate.
Practical implementations could be adapted from proposals of quantum switch and interferometric based gates~\cite{Micheli2004, Mueller2009, Abanin2012}, or schemes for work distribution measurements~\cite{Dorner2013,Mazzola2013}.
Notice that if the Hamiltonian $-H$ is also applicable on the system~\cite{Swingle2016}, one may also access the overlap from $\frac{1}{\sqrt{2}}(\ket{\psi(-t/2)}\ket{0}_{\text{p}} + \ket{\psi(t/2)}\ket{1}_{\text{p}})$.Using the purification scheme, a strategy experimentally used in Ref.~\cite{Islam2015}, $G(t)$ could be measured by starting the interferometric protocol from $\ket{\psi_{\infty}}$ and applying the prescribed time evolution on the $S$ part only. An expensive alternative without ancillas is to start from $\ket{\psi_0}=\ket{\sigma}$ to measure 
\begin{equation}
G_\sigma(t) = \bra{\sigma}e^{-it H}\ket{\sigma} = \sum_n |{\braket{\sigma}{n}}|^2 e^{-itE_n}\;,
\label{eq:Gsigma}
\end{equation}
and collect all $G_\sigma(t)$ to reconstruct  $G(t)=\frac 1 D \sum_\sigma G_\sigma(t)$.
Last, in analogy to finite-temperature Lanczos approaches~\cite{Prelovsek2013,Steinigeweg2014} and for large enough Hilbert spaces, the trace in $G(t)$ could be approximated by sampling over random states instead of Fock states. 
Numerically, $G(t) = \braket{\psi_{\infty}}{\psi(t)}$ is accessible through time propagation of MPS, similarly to finite-temperature calculations~\cite{Verstraete2004,Feiguin2005, Barthel2009}. This allows for large system estimates in one dimension as shown in the supplemental material~\cite{Sup} using the Itensor library~\cite{Itensor}. 
In the following, we simply use full diagonalization of small systems to illustrate the information one can retrieve from such purification spectroscopy.
In the context of dynamical quantum phase transitions, studying $Z(it)$ has the advantage of discussing the occurrences of non-analyticities through a quantity that is independent of the initial state $\ket{\psi_0}$. 

\begin{figure}[t]%
\includegraphics[width=\columnwidth]{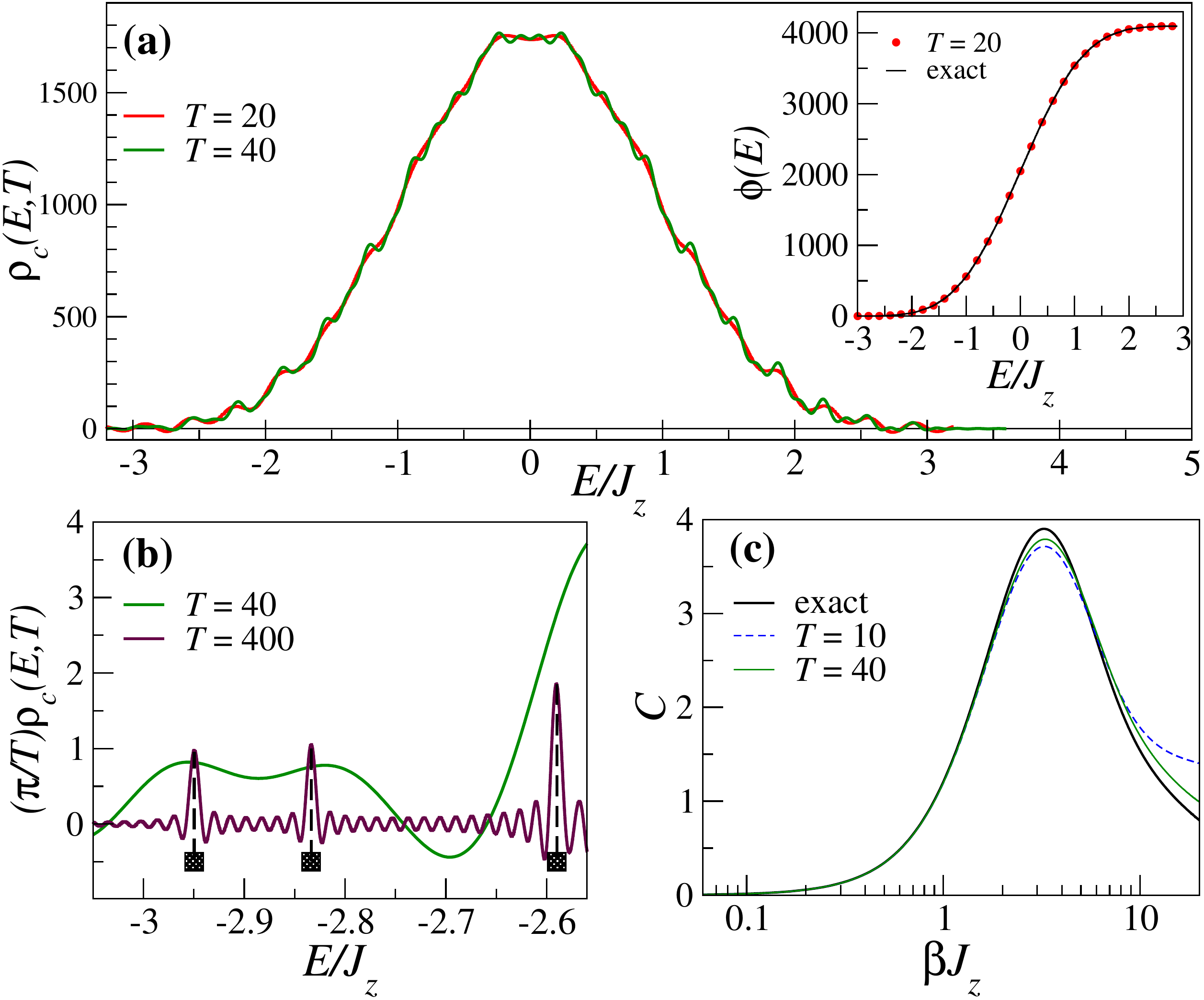}%
\caption{Illustration of the reconstructed density of state $\rho_c(E,T)$ of Hamiltonian \eqref{eq:HamIsing} for $L=12$, $h_x=0.2J_z$ and $h_z=0.01J_z$. 
Times $T$ given in units of $J_z^{-1}$.
\textbf{(a)} Coarse-grained density of states $\rho_c(E,T)$. Inset: integrated density of states $\phi(E) = \int_{-\infty}^E dE'\rho_c(E',T)$ compared to the exact one.
\textbf{(b)} Zoom on the first low-energy states showing two near degenerate ground-states and the gap to the first (doubly degenerated) excitation.
\textbf{(c)} Reconstruction of the specific heat $C(\beta)$ from Eq.~\eqref{eq:thermo} for $h_x=0.5J_z$.}
\label{fig:rho_spectro}%
\end{figure}

\textbf{Density of states} -- By definition, $Z(it)$ is the Fourier transform of the density of states $\rho(E) = \sum_n \delta(E-E_n)$. Measuring $G(t)$ up to time $T$allows to construct the coarse-grained density of states
\begin{equation}
\rho_c(E,T) = \int_{0}^{T} \!\frac{dt}{\pi} \!\Re{Z(it)e^{itE}} = \sum_n \delta_T(E-E_n)\,,
\end{equation}
in which $\delta_T(\varepsilon)=\sin(T\varepsilon)/(\pi\varepsilon)$ is a normalized sinc function of typical width $1/T$, and with $\delta_T(0) = T/\pi$. We thus have
\begin{equation}
\lim_{T\to\infty}\rho_c(E,T) = \rho(E)\;.
\end{equation}
The evolution of $\rho_c(E,T)$ with time $T$ is governed by the comparison between the width $1/T$ of the $\delta_T$ peaks and the local level spacing $1/\rho(E)$.
At very short time $\rho_c(E,T)\simeq \frac{T}{\pi}D$. 
We then define a typical time $T_c(E) = \rho(E)$ such that for $T\ll T_c(E)$, $\rho_c(E,T)$ builds up a local coarse-grained representation of $\rho(E)$ and for $T\gg T_c(E)$, $\rho_c(E,T)$ resolves the local energy peaks.
At the edges of the spectrum, $T_c(E)$ is either finite (for a gaped system) or scales polynomially with the system size.
In the bulk of the spectrum, $T_c(E)$ is typically exponential in the system size.
These behaviors are illustrated in Fig~\ref{fig:rho_spectro}(a-b) on a system with $L=12$ with Hamiltonian~\eqref{eq:HamIsing}.
For relatively short times $T\sim 20 J_z^{-1}$, the overall gaussian shape~\cite{Atas2014} of the density of states is well reproduced (Fig~\ref{fig:rho_spectro}(a)).
As $T$ increases, one observes the low-energy peaks corresponding to the near-degenerate ground-states and the gap to the first excitation state (Fig~\ref{fig:rho_spectro}(b)).
Experimentally, observing both quantities would be remarkable, already on small systems. 
In numerics, only the coarse-grained approximation of $\rho(E)$ would be interesting, since taking the inverse Fourier transform is less sensitive to noise than inverse Laplace transform that one would perform from finite temperature estimations of the partition function $Z(\beta)$.

\textbf{Semi-classical reconstruction of thermodynamics} -- A possible application is to use the coarse-grained functions to reconstruct averages over the spectrum. 
For an observable ${A}$, we define the reconstructed statistical average as
\begin{equation}
\label{eq:thermo}
\overline{A}(\beta;T) = \int_{E_0}^{E_{\text{max}}}dE\,\rho_c(E,T)\,A_c(E,T)\,\frac{e^{-\beta E}}{Z_T(\beta)}
\end{equation}
in which $\beta$ is the inverse temperature, $E_0(E_{\text{max}})$ are estimates of the lowest(highest) energies, $Z_T(\beta) = \int_{E_0}^{E_{\text{max}}}dE\,\rho_c(E,T)\,e^{-\beta E}$ and $A_c(E,T)$ a coarse-grained estimate of the $A_n$ (see the paragraph on observables spectroscopy below), or a function of the energy $E$.
For instance, we show on Fig~\ref{fig:rho_spectro}(c) the reconstruction of the specific heat $C = \beta^2(\overline{E^2}(\beta;T)-\overline{E(\beta;T)}^2)$ of the model as a function of $\beta$, for various time $T$.
Thus, even from relativity short times $T$, one retrieves relevant thermodynamical observables from studying the time-evolution of a quantum simulator.

\begin{figure}[t]%
\includegraphics[width=\columnwidth]{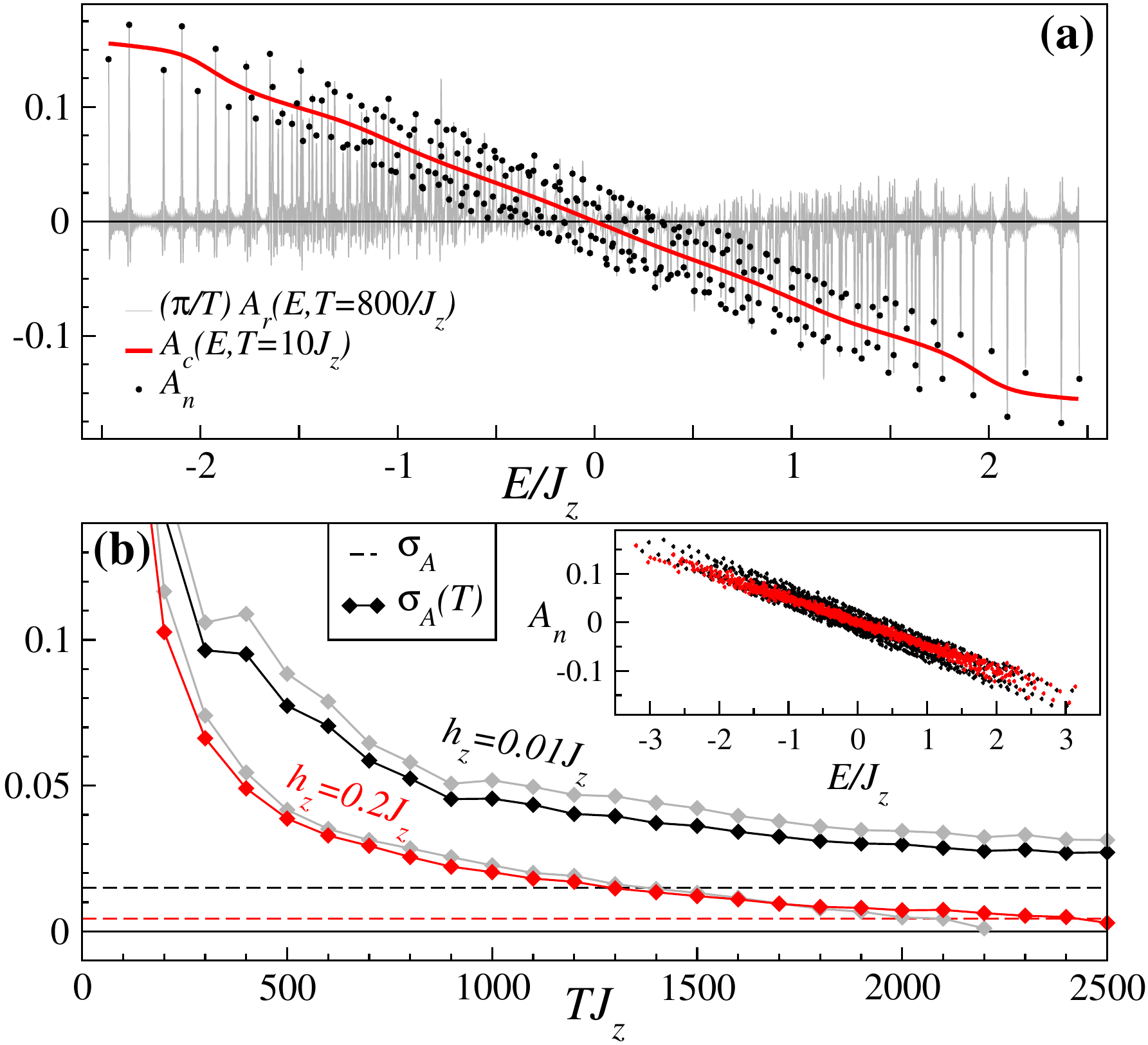}
\caption{Observable spectroscopy using $A=\frac 1 {L-1}\sum_{\ell=1}^{L-1} S^z_\ell S^z_{\ell+1}$.
\textbf{(a)} Typical behavior of the functions $A_c(E,T)$ and $\frac{\pi}{T}A_r(E,T)$ compared to the exact $A_n$ for a system with $L=8$, $h_x=0.5J_z$ and $h_z=0.01J_z$.
\textbf{(b)} Fluctuations $\sigma_A(T)$ as a function of $T$ for $L=10$, $T_{sc}J_z=10$, $h_x=0.5J_z$ and two values $h_z=0.01J_z$ (near integrable)  $h_z=0.2J_z$ (non-integrable) using an energy range $E_{\pm} = \pm J_z$ (and $E_{\pm} = \pm 1.3J_z$ for gray curves). Dotted lines show the expected limit $\sigma_A$. 
Inset: exact $A_n$ (same color code).}
\label{fig:Aspectro}
\end{figure}

\textbf{Observables spectroscopy} -- The spectroscopy of the eigenstates expectations $A_n$ for an observable ${A}$ requires to apply $A$ at the end of time evolution.
For simplicity, one may think of a diagonal operator in the Fock basis.
Indeed, we define for the $S\otimes \tilde{S}$ full system
\begin{align}
G_A(t) &= \bra{\psi_{\infty}}{A}\ket{\psi(t)} =
\frac{1}{D} \Tr({A}e^{-it H}) = \frac{A(t)}{D}\,,
\end{align}
in which we define, for $t>0$, $A(t) = \sum_{n} A_n e^{-it E_n}$, the Fourier transform of peaks of magnitude $A_n$ located at positions $E_n$ in energy.
After defining 
\begin{equation*}
A_r(E,T) = \int_{0}^{T} \!\frac{dt}{\pi} \!\Re{A(t)e^{itE}}= \sum_n A_n \delta_T(E-E_n)\,,
\end{equation*}
the coarse-grained function for observable $\hat{A}$ reads
\begin{equation}
A_c(E,T) = \frac{A_r(E,T)}{\rho_c(E,T)} \;.
\end{equation}
The interpretation of $A_c$ is transparent: it averages over a small window of energy $1/T$ the local contributions of the $A_n$.
The typical time scales are the same as for the density of states.
At some intermediate times $T\ll T_c(E)$, $A_c(E,T)$ provides a smooth function of the energy, which one could compare to semi-classical predictions~\cite{Tomsovic2017}. On small systems, one resolves the $A_n$ from
\begin{equation}
\lim_{T\to\infty} \frac{\pi}{T}A_r(E_n,T) = A_n\;.
\end{equation}
We stress that this protocol allows to access the coarse-grained function, without theoretical input, either from numerics or from experiments being able to measure $G_A(t)$.
We illustrate the typical behavior of the $A_r(E,T)$ and $A_c(E,T)$ functions for various $T$ in Fig.~\ref{fig:Aspectro}(a).
A short intermediate times, the coarse-grained function $A_c(E,T)$ could be used as input to predict thermodynamical averages as Eq.~\eqref{eq:thermo}.
We now turn to the possibility, using long-time observations, to discuss how the $A_n$ fluctuate around this semi-classical prediction.

\textbf{Towards experimental tests of the ETH} --
Interestingly, comparing the large $T$ and small $T$ data, on a finite system, makes it possible to test the ETH. 
Basically, the ETH works when there is an equivalence of ensembles and if one can replace the $A_n$ by a coarse grained function $A_c(E_n)$~\cite{Jancel1969,Peres1984,Deutsch1991,Srednicki1994,Tasaki1998,Srednicki1999,Rigol2008,Reimann2008,Roux2009,Roux2010a,Roux2010,Biroli2010,DAlessio2015,Borgonovi2016}, as one usually does in statistical physics. 
Equivalently, this means that the fluctuations of $A_n$ will yield a negligible contribution in the thermodynamical limit, for which numerical tests have been carried out~\cite{Jensen1985,Feingold1986,Ikeda2013,Steinigeweg2014,Beugeling2014}.
The coarse-grained function $A_c(E_n,T_{sc})$, with $T_{sc} \ll T_c(E)$~\cite{Sup}, allows to extract the fluctuations $A_n - A_c(E_n,T_{sc})$ and define 
\begin{equation}
\label{eq:sigmaA}
\sigma_A^2 = \frac{1}{D_{\mathcal{N}}}\sum_{n \in \mathcal{N}} (A_n - A_c(E_n,T_{sc}))^2\;,
\end{equation}
in which $\mathcal{N}$ is the subset of eigenstates, of size $D_{\mathcal{N}}$, belonging to some energy range $[E_-,E_+]$ (typically in the bulk of the spectrum).
From signals of observable spectroscopy, we may qualitatively access such fluctuations using
\begin{equation*}
\sigma_A^2(T) = \frac{\int_{E_-}^{E_+}\!\!dE \left(\frac{\pi}{T}A_r(E,T)-A_c(E,T_{sc})\right)^2\rho_c(E,T)}{\int_{E_-}^{E_+}\!\!dE \rho_c(E,T_{sc})}\;,
\end{equation*}
and look for long times $T$.
We illustrate this approach in Fig.~\ref{fig:Aspectro}(b) where the fluctuations $\sigma_A(T)$ are plotted as a function of $T$ for near-integrable ($h_z=0.01J_z$) and non-integrable regimes ($h_z=0.2J_z$). Such integrals are numerically non-trivial because of the $\delta_T$, and the convergence towards the expected $\sigma_A$ limit requires times of the order of $T_c$, is rather slow and slightly depends on the choice of $E_{\pm}$. Nevertheless, one already observes a qualitative difference between the two regimes at finite time $T$.

\begin{figure}[t]
\includegraphics[width=\columnwidth]{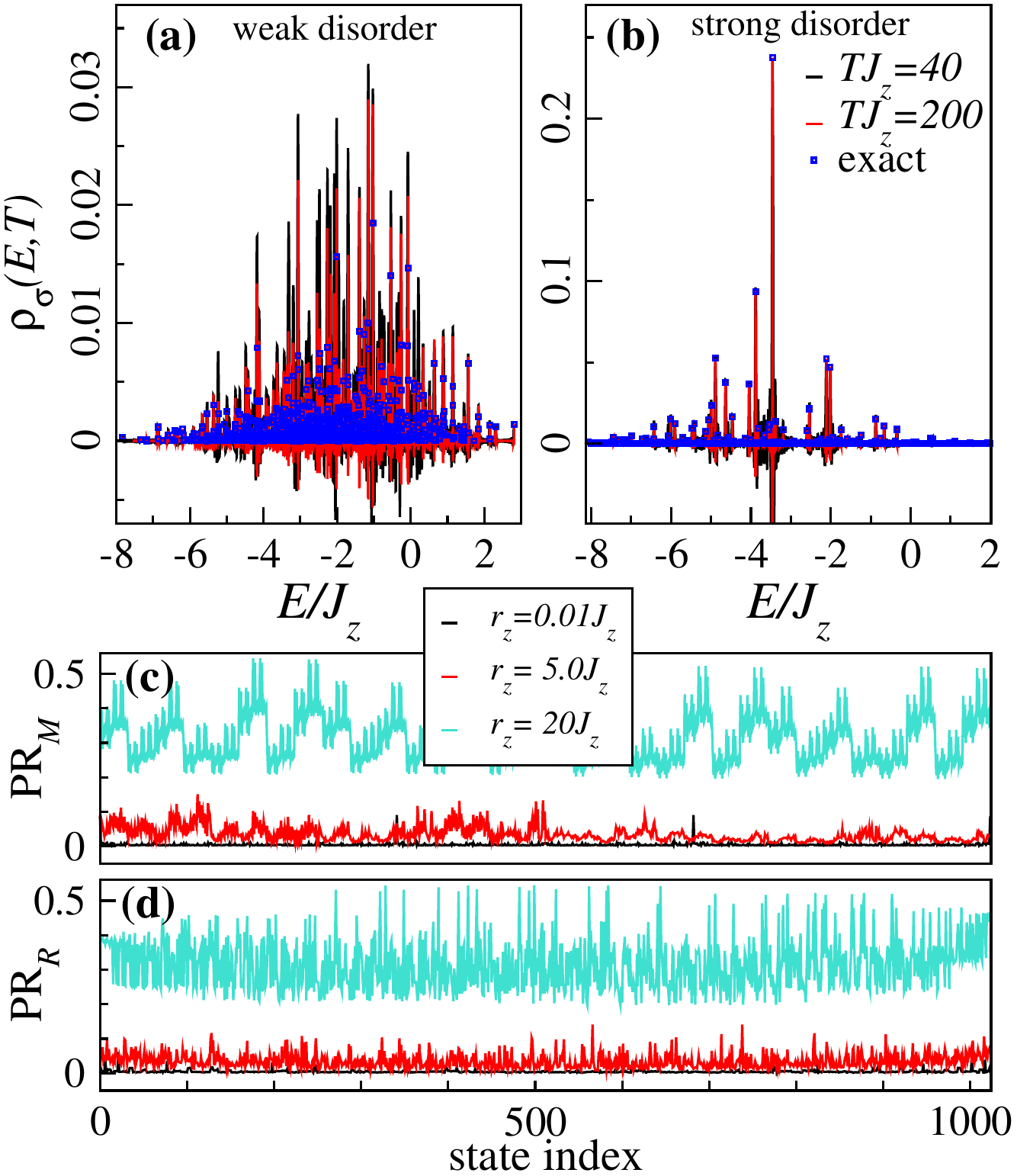}
\caption{\textbf{(a-b)} Illustration of (bulk) Fock state energy distribution $\rho_\sigma(E,T)$ on a $L=10$ chain with $\ket{\sigma} = \ket{\uparrow\uparrow\downarrow\downarrow\uparrow\uparrow\downarrow\downarrow\uparrow\uparrow}$, starting from the paramagnetic regime $h_x=J_z$.
The energy localization is seen at large disorder $r_z=5J_z$ \textbf{(b)} while energy delocalization is found at low disorder $r_z=0.1J_z$ \textbf{(a)}.
\textbf{(c-d)} Participation ratios of the columns of $M$ and $R$ matrices for increasing disorder $r_z$.}
\label{fig:MBL}
\end{figure}

\textbf{Fock state spectroscopy and many-body localization} -- 
We previously discussed the possibility to measure $G_\sigma(t)$ using the interferometric protocol.
Eq.~\eqref{eq:Gsigma} directly shows that it would offer, by Fourier transform, access to the distribution of the weights $|{\braket{\sigma}{n}}|^2$ describing how $\ket{\sigma}$ is spread over the eigenstates.
Consequently, we introduce the matrix $M$ with entries $M_{\sigma n} =|{\braket{\sigma}{n}}|^2$ that is real positive, bistochastic and non-symmetric.
At time $T$, it provides a coarse-grained reconstruction $\rho_\sigma(E,T) = \frac{\pi}{T}\sum_n M_{\sigma n} \delta_T(E-E_n)$ of the Fock state energy distribution.
As a practical application, we show on Fig.~\ref{fig:MBL}(a-b) the typical behavior of $\rho_\sigma(E,T)$ in the situation of many-body localization, by adding a disorder term $\sum_\ell h_\ell S^z_\ell$ in \eqref{eq:HamIsing}, with random longitudinal fields $h_\ell$ uniformly distributed in the range $[-r_z/2,r_z/2]$ (see also~\cite{Sup} for MPS calculations).
Starting from the paramagnetic phase in which a typical bulk Fock state is naturally delocalized over the eigenstates (Fig.~\ref{fig:MBL}(a)), increasing the disorder makes the Fock state closer and closer to an exact eigenstate, building a localized energy distribution (Fig.~\ref{fig:MBL}(b)).
This approach is the reciprocal idea of the localization of eigenstates along the Fock states, or related quantities~\cite{Basko2006,Pal2010,Murphy2011,Serbyn2013,Luca2013,Luitz2015}, and related to typical experimental protocols starting from Fock states~\cite{Schreiber2015}. Interestingly, this protocol extends to the many-body case the measurements of the $M_{\sigma n}$ coefficients recently performed for single and two-particle states using a different spectroscopy technique~\cite{Roushan2017}.
In order to get an overall spectral picture of the localization of Fock states, we show in Fig.~\ref{fig:MBL}(c) the participation ratio $\text{PR}_{M}(\sigma)=\sum_n M_{\sigma n}^2$ as a function of index $\sigma$. Localization globally occurs at large disorder when $\text{PR}_{M}(\sigma)$ is of order one.

Last, we notice that one cannot access $M$ without the interferometric protocol, by measuring probabilities only, ie. from measurements that are invariant under the global phase of $\ket{\psi(t)}$.
One rather accesses to a matrix corresponding to Uhlmann amplitudes~\cite{Uhlmann1986,Uhlmann1995}.
First consider the time evolution of the purified state density-matrix and measure the density-matrix elements 
\begin{equation}
\Gamma_{\sigma\sigma'}(t) = \braket{\sigma,\sigma}{\psi(t)}\braket{\psi(t)}{\sigma',\sigma'} = \frac{1}{D} M C(t) M^\mathsf{T}
\label{eq:GammaMatrix}
\end{equation}
in which we define the $C$ matrix as $C_{nn'}(t) = e^{-it(E_n-E_{n'})}$.
After averaging over time $\overline{\Gamma}(T) = \frac{1}{T}\int_0^T dt \Re{\Gamma(t)} = \frac{1}{D}MS(T)M^\mathsf{T}$, with $S_{nn'} = \frac{\pi}{T}\delta_T(E_n-E_{n'})$. 
Assuming a spectrum without degeneracies, we have in the long time limit $\overline{\Gamma}(\infty) = \frac{1}{D}MM^\mathsf{T}$, a positive symmetric matrix.
Due to gauge invariance, one cannot retrieve $M$ from $\overline{\Gamma}(\infty)$ but one has access to the matrix $R = \sqrt{D\overline{\Gamma}(\infty)}$.
Indeed, the polar decomposition $M=R U$ of $M$ is unique, with $R=\sqrt{MM^\mathsf{T}}$ the symmetric positive definite matrix containing the Uhlmann amplitudes and $U$ a unitary matrix.
Interestingly, we show in Fig.~\ref{fig:MBL}(d) that the participation ratio $\text{PR}_{R}(\sigma)$ of the columns of $R$ also displays a delocalization-localization crossover with increasing disorder.
At finite time $T$, provided the positiveness of $\overline{\Gamma}(T)$, one has a semi-classical representation $R(T) = \sqrt{D\overline{\Gamma}(T)}$.
Alternatively, averaging over time the probabilities $p_{\sigma,\sigma'}(t) = |{\braket{\sigma,\sigma'}{\psi(t)}}|^2$ in the purification scheme, with
\begin{equation*}
p_{\sigma,\sigma'}(t)
= \frac{1}{D} \sum_{n,n'}\braket{\sigma}{n}\braket{n}{\sigma'}\braket{n'}{\sigma}\braket{\sigma'}{n'}C_{nn'}(t)\,,
\end{equation*}
gives back the $\overline{\Gamma}(\infty)$ matrix elements~\footnote{Without any purification scheme but measuring the probabilities $\tilde{p}_{\sigma,\sigma'}(t)=|{\bra{\sigma'}e^{-iHt}\ket{\sigma}}|^2$ directly on the system, by scanning all initial and final Fock states gives after time averaging $\overline{\tilde{p}}(\infty) = MM^{\mathsf{T}}$.}.

In conclusion, we propose a protocol to experimentally access, on small systems, the essential spectral features of an Hamiltonian in a coarse-grained fashion.
Such protocol could be a route for testing more directly the mechanisms behind ETH and MBL.
Resolving the eigenstates features would require long lifetime quantum simulator, for instance envisioned in Ref.~\cite{Nguyen2018}, but we see that intermediate times already bring relevant coarse-grained quantities for the model, which could be compared to theoretical predictions.

We thank F. Alet, A. Browayes, A. del Campo, M. Heyl, K. Le Hur, O. Giraud, A. Polkovnikov, C. Sayrin and O. Viyuela for stimulating discussions.
We acknowledge funding by the ANR under the project TRYAQS (ANR-16-CE30-0026).

\bibliographystyle{apsrev4-1}
\bibliography{DynamicalPhaseTransition}

\cleardoublepage
\onecolumngrid
\appendix

\begin{center}
{\Large\textbf{Supplementary material for: Quantum purification spectroscopy}}
\end{center}

\section{Typical time scales in the $A_c(E,T)$ function}

\begin{figure}[h]
\includegraphics[width=10cm]{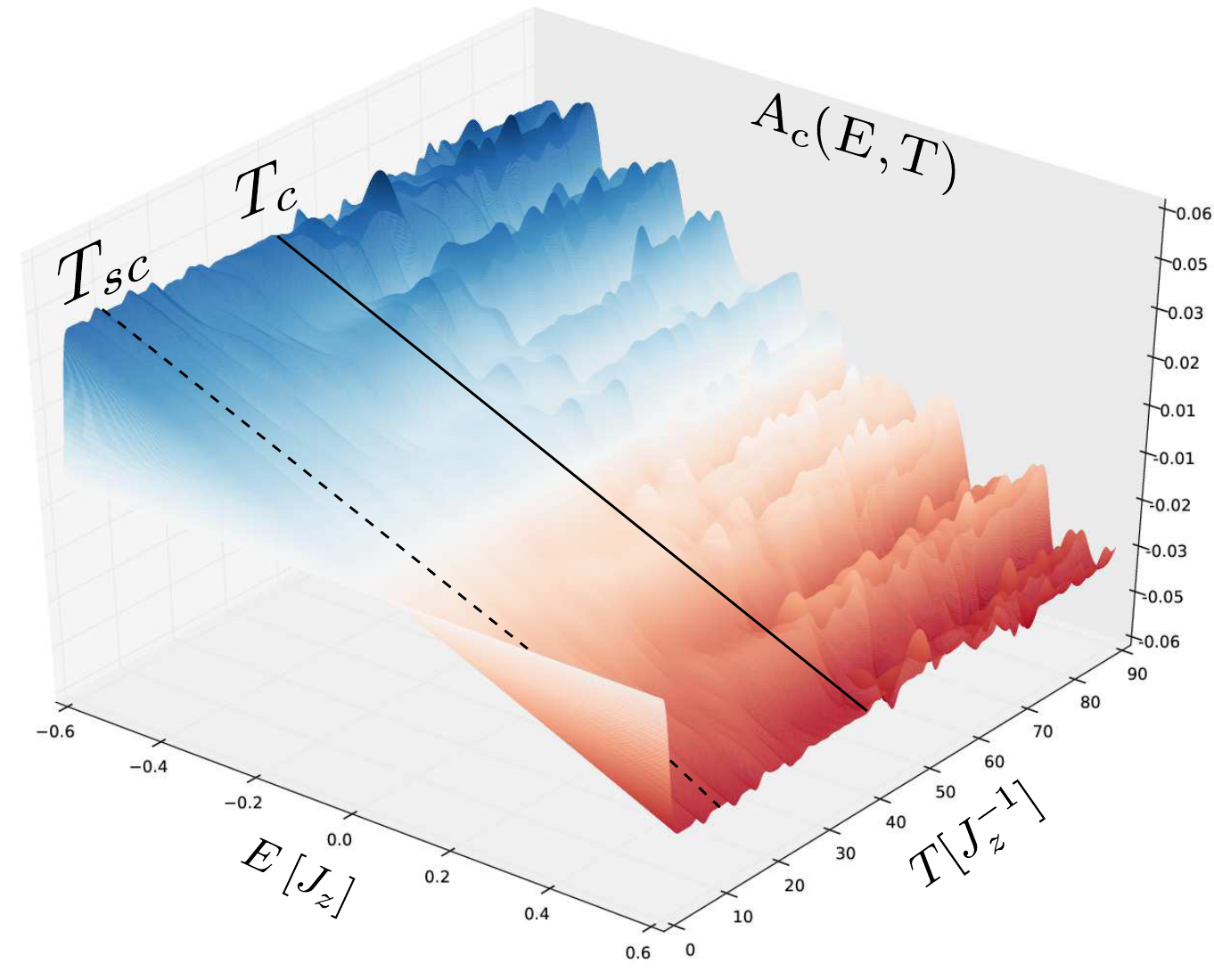}
\caption{Typical behavior of the $A_c(E,T)$ function in a small window of energy so as to make the behavior clearer. Parameters are the same as in Fig.~\ref{fig:Aspectro}(a). Here $T_{sc}J_z = 10$.}
\label{fig:supAc}
\end{figure}

We here give in Fig.~\ref{fig:supAc} the typical behavior of the coarse-grained function $A_c(E,T)$ as a function of observational time $T$ in order to explain the meaning of the $T_{\text{sc}}$ and $T_c$ times. The model and observable $A$ are the same as in the main text and Fig.~\ref{fig:Aspectro}.

As discussed in the main text, $T_c$ is the typical time where the width of the $\delta_T$ functions are of the same order of magnitude as the local level spacings. This motivates us to take the definition $T_c(E) = \rho(E)$ but one can keep in mind that $T_c$ is a crossover time.
In Fig.~\ref{fig:supAc}, we clearly observe a smooth behavior for $T<T_c$ and peaks starting to emerge for $T>T_c$ in this window of energy. The smooth (almost linear in the window of energy) behavior of $A_c(E,T)$ for $T<T_c$ corresponds to the coarse-grained regime where the $\delta_T$ functions perform a local average of the eigenstates contributions. Thus, one can use a typical time $T_{sc}<T_c$, and in practice $T_{sc}\ll T_c$ for large systems, to define a coarse-grained prediction $A_c(E,T_{sc})$ for the behavior of the observable as a function of energy. There is no need for a precise definition of $T_{sc}$, on just requires it to be in the coarse grained regime. Notice that at very low times $T$, $A_c(E,T)$ is a constant as it integrates contributions over the whole spectrum.

\section{Examples of calculations using Matrix Product States algorithms}

The goal of this section is to provide a proof of concepts example using Matrix Product States (MPS) calculations.
These numerical simulations are carried out using the Itensor library~\cite{Itensor} to implement the protocol.

\subsection{Density of states}

First, using the purification protocol, we show an example of coarse-grained density of states obtained for another Hamiltonian than the one discussed in the main text.
We choose the XXZ model, written as
\begin{equation}
\label{eq:HamXXZ}
H = \sum_{\ell=1}^{L-1} \left[ J_z S^z_{\ell}S^z_{\ell + 1} + J( S^x_{\ell} S^x_{\ell+1} + S^y_{\ell} S^y_{\ell+1})\right],
\end{equation}
and open boundary conditions for the following reasons: open boundary conditions are more suitable for MPS calculations, the model conserves the total spin along the $z$ direction, making the Hamiltonian block-diagonal, and its density of states is not easy to capture since the exact spectrum can be computed by Bethe ansatz but large systems become tedious to handle.
Yet, in the large $J_z$ limit dominiated by the Ising limit of the model, we expect to recover the multi-gaussian peaks features discussed in Ref.~\onlinecite{Atas2014} on the exactly solvable Ising in transverse field model. Thus, we show a practical example for a large system $L=30$, which goes beyond other kinds of methods (in particular full diagonalization).

\begin{figure}[h]
\includegraphics[width=0.8\columnwidth]{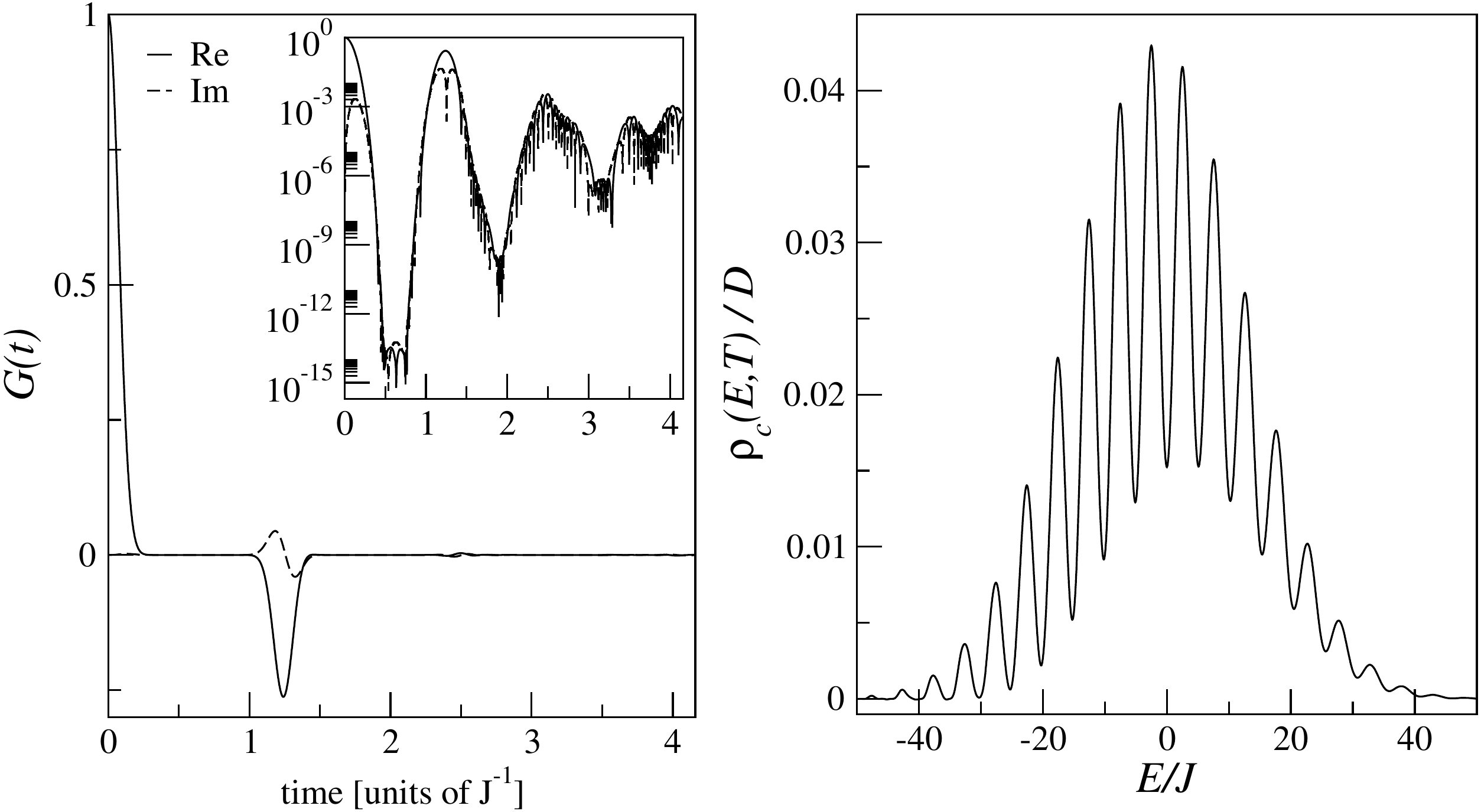}
\caption{Example of an MPS calculation of the $G(t)$ function for Hamiltonian~\eqref{eq:HamXXZ} with $J_z=10J$, $L=30$ and open boundary conditions. The left panel displays the real and imaginary parts of the $G(t)$ function (the inset shows their absolute value in log-scale), while the right panel shows the coarse-grained density of states $\rho_c(E,T)$ computed for $T\simeq 4.16 J$. The main peaks are reminiscent of the Ising limit and are not artefacts of the finite $T$ pseudo-Fourier transform. }
\label{fig:supDOS}
\end{figure}

Results for $J_z=10J$ are reported in Fig.~\ref{fig:supDOS} and show that getting the $G(t)$ function for a moderate time $T\simeq 4J^{-1}$ already allows to reconstruct non-trivial features of the density of states. The main peaks are reminiscent of the Ising limit where typical excitations are separated by $J_z/2 = 5 J$ gaps. These peaks are broadened by the transverse quantum fluctuations associated with $J$, with an overall behavior qualitatively similar to the ones observed in Ref.~\onlinecite{Atas2014}.

\subsection{Fock state spectroscopy for many-body localization}

We now turn to the calculation of $G_\sigma(t)$ functions to probe the spectral decomposition of Fock states using MPS calculations. We use the Ising in transverse field model with random longitudinal fields as presented in the main text for systems with $L=20$ and $L=30$. The Fock state is the same as in the main text and lies in the bulk of the spectrum. 

\begin{figure}[h]
\includegraphics[width=\columnwidth]{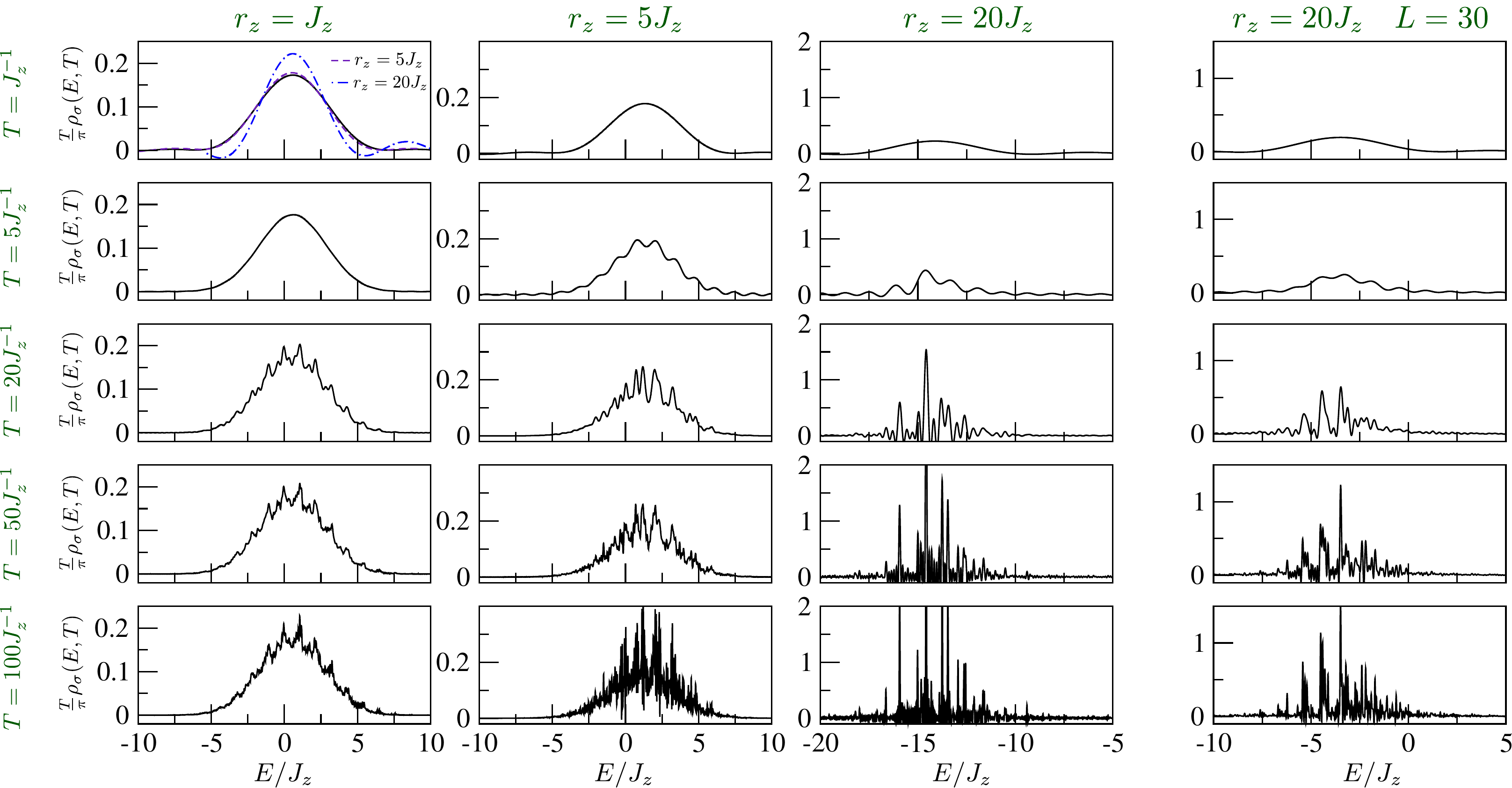}
\caption{Example of $\rho_\sigma(E,T)$ functions for increasing $T$ and disorder $r_z$. The three leftmost columns are on a system with $L=20$. The rightmost column is for $L=30$. The total observation time $T$ increases from top to bottom. On the top left panel, the three $L=20$ results for different disorder are gathered in the same plot (with a shift in energy to put them on each other) to show that they qualitatively look the same at short $T$.}
\label{fig:supMBL}
\end{figure}

The results for the coarse-grained function $\rho_\sigma(E,T)$ are gathered in Fig.~\ref{fig:supMBL} for increasing $T$ and disorder strength $r_z$.
Interestingly, for a short observation time such as $T=J_z^{-1}$, a gaussian like spectrum is obtained in all cases, but increasing $T$ up to $100J_z^{-1}$ shows that the distribution gets decomposed into sharped peaks as the many-body localized regime is reached. These calculations are proof of concepts and not meant to be a detailed analysis of the many-body localization in this model, something beyond the scope of this manuscript.

\begin{figure}[h]
\includegraphics[width=10cm]{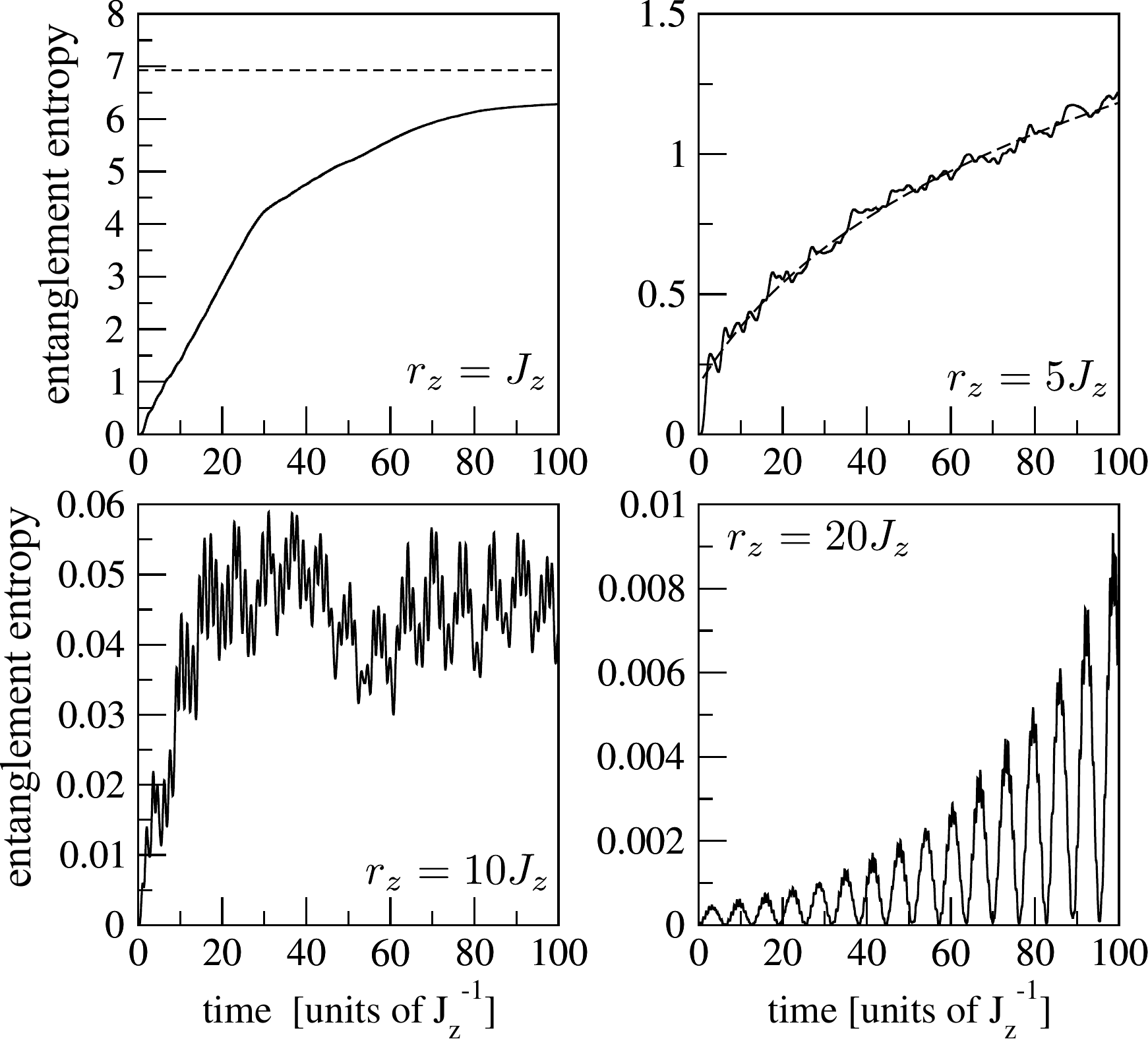}
\caption{Entanglement entropy as a function of time for the $L=20$ data of Fig.~\ref{fig:supMBL}. 
In the top left panel, the dashed line corresponds to the maximum possible entropy $\frac{L}{2}\ln 2$.
In the top right panel, the dashed line corresponds to a logarithmic fit of the evolution.}
\label{fig:supMBLsvn}
\end{figure}

As expected, the calculation time is strongly dependent on the physics of the system. 
We illustrate this aspect by showing in Fig.~\ref{fig:supMBLsvn} the evolution of the entanglement entropy in the $L=20$ calculations discussed above.
The entanglement entropy is taken to be the von Neumann entropy of half of the system.
In the delocalized regime, it typically grows linearly in time. In this example, it even reaches a value close to the maximum possible entropy showing that the state is delocalized almost over the whole Hilbert space. Such calculations in this regime are hard on long times, as in similar time-dependent studies of many-body delocalization. 
When disorder is increased, we observe a typical logarithmic behavior~\cite{Znidaric2008} of the entanglement entropy, while at very large disorder and on the time scales that are explored, the entropy remains very small, which allows to perform long time simulations that are fast.

\end{document}